\documentclass[aps,prb,preprint,showpacs]{revtex4}

\usepackage{epsfig}

\begin{document}
\newcommand{\vektor}[1]{\mbox{\boldmath $#1$}}
\newcommand{\eff}{\mbox{\scriptsize{eff}}}
\newcommand{\Hub}{\mbox{\scriptsize{Hub}}}
\newcommand{\Hol}{\mbox{\scriptsize{Holstein}}}
\newcommand{\SSH}{\mbox{\scriptsize{SSH}}}
\newcommand{\crit}{\mbox{\scriptsize{crit}}}
\newcommand{\doex}{\mbox{\scriptsize{DE}}}
\newcommand{\KL}{\mbox{\scriptsize{KL}}}
\newcommand{\KLph}{\mbox{\scriptsize{PKL}}}
\newcommand{\DKL}{\mbox{\scriptsize{DKL}}}
\newcommand{\AKL}{\mbox{\scriptsize{AKL}}}
\newcommand{\RKKY}{\mbox{\scriptsize{RKKY}}}
\newcommand{\PAM}{\mbox{\scriptsize{PAM}}}
\title{Spin and lattice effects in the Kondo lattice model} 
\author{M. Gulacsi${}^{\rm 1}$, A. Bussmann-Holder${}^{\rm 2}$ 
and A. R. Bishop${}^{\rm 3}$}
\affiliation{
${}^{\rm 1}$ 
Department of Theoretical Physics, 
Institute of Advanced Studies, 
The Australian National University, Canberra, ACT 0200, Australia \\
${}^{\rm 2}$ 
Max-Planck-Institut f\"ur Festk\"orperforschung 
Heisenbergstr. 1, 70569 Stuttgart, Germany 
${}^{\rm 3}$ 
Theoretical Division, Los Alamos National Laboratory, 
Los Alamos, NM 87545, U.S.A.}

\date{\today}

\begin{abstract}
The magnetic properties of a system of coexisting localized 
spins and conduction electrons are investigated within an 
extended version of the one dimensional Kondo lattice model
in which effects stemming from the electron-lattice and  
on-site Coulomb interactions are explicitly included. After 
bosonizing the conduction electrons, is it observed that 
intrinsic inhomogeneities with the statistical scaling 
properties of a Griffiths phase appear, and determine 
the spin structure of the localized impurities. The 
appearance of the inhomogeneities is enhanced by appropriate
phonons and acts destructively on the spin ordering. The 
inhomogeneities appear on well defined length scales and
can be compared to the formation of intrinsic mesoscopic 
metastable patterns which are found in two-fluid
phenomenologies. 
\end{abstract}

\maketitle

\section{Introduction}

The interplay of spin, charge and lattice degrees of freedom 
has been investigated intensively in many transition metal 
oxides and especially in perovskite manganites, which 
have recently attracted new interest due to the discovery of 
colossal magnetoresistance (CMR). The initial understanding
of the properties of manganites was based on the double-exchange
mechanism within a Kondo lattice.\cite{deCMR} However, 
neutron scattering \cite{hayden} and electron diffraction \cite{chen} 
experiments have revealed the simultaneous presence of charge and
spin superstructures which makes these early theoretical approaches 
incomplete. 

The existence of charge, lattice and spin modulations with a 
doping-dependent wave vector in La${}_{2-x}$Sr${}_{x}$NiO${}_{4}$,
\cite{hayden,chen} or the presence of charge ordering at half filling
in Nd${}_{0.5}$Sr${}_{0.5}$MnO${}_{3}$ \cite{kawano} and similar 
compounds, such as Pr${}_{0.5}$Ca${}_{0.5}$MnO${}_{3}$,
\cite{tomioka,li,moritomo,chen2} 
all suggest that the theoretical understanding has to be extended 
to account for effects stemming from the lattice in order to 
understand the doping dependent phase diagram and the 
richness of phases that are obtained. 
More recently, in electron doped charge ordered manganites 
La${}_{1-x}$Ca${}_{x}$MnO${}_{3}$ charge, orbital, and magnetic 
ordering has been observed \cite{wollan} for the first time. 
Several experiments also confirmed this behavior in other 
compounds.\cite{liubao,mori,ueharajirac} 
Charge ordering has also been found for other doping systems as 
in Bi${}_{1-x}$Ca${}_{x}$MnO${}_{3}$ \cite{liubao} 
and in La${}_{1-x}$Ca${}_{x}$MnO${}_{3}$  
(doped with $P_{r}$) \cite{ueharajirac} for $x \geq 1 / 2$. 

Because of these complex behaviors, manganese 
oxides have been intensively studied during the 
last few years. Since realistic models of these perovskites 
is impossible to solve completely, different approximation schemes
have been introduced to account for the individual 
properties of their rich phase diagram. Typically, the
electronic degrees of freedom are described by a Kondo lattice
model (KLM) which in the strong Hund coupling limit reduces to the Zener
\cite{zener51} double-exchange hamiltonian. Early theories \cite{deCMR} 
proposed to explain
the physics of CMR materials have focused primarily on this model.
As the previously noted experimental findings show, all the
approaches based entirely on the phenomenon of double-exchange 
are incomplete \cite{phonontheory}. Understanding the complex
phase diagram of the perovskites can be resolved by including 
additional physics in the Kondo lattice, in the form of electron-phonon 
coupling \cite{phonontheory} originating from a Jahn-Teller
splitting of the Mn${}^{3+}$ ion. Consequently, our first goal here is
to understand the lattice effects in the KLM by including explicitly 
the interaction with the lattice degrees of freedom.

A further consequence of the strong Hund coupling is that
three localized $t^3_{2g}$ orbitals of the 
Mn${}^{4+}$ ions will be aligned, giving rise to a $S = 3/2$ localized spin. 
As most of the recent approaches to CMR are based on Monte Carlo simulations 
of one dimensional models,\cite{MC} it has been argued that for finite 
temperatures a classical spin represents a reasonable approximation. 
However, this may not be the best approach, as the quantum fluctuations
are the strongest in one dimension. Hence, our second goal here is
to develop a bosonization approach which solves the full quantum 
spin KLM in one dimension.
For completeness we will allow for both ferro- and antiferromagnetic
coupling in the KLM. In CMR materials where the coupling is
ferromagnetic, it is obvious that double-exchange dominates. We will
show in Section II that this is also the case for antiferromagnetic 
coupling, as in, e.g., heavy fermion compounds \cite{heavyfermion}. 

The paper is organised as follows, in Section II we investigate
the presence of double-exchange ferromagnetism in both $J < 0$
and $J > 0$ cases. Section III contains a description of  
our starting hamiltonian. Section IV contains the details of the
bosonized solution. Section V is devoted to a comprehensive 
description of the localized spin ordering. In Section VI the
obtained phase diagram is analysed in detail. Section VII 
contains the conclusions. 

\section{Double-exchange}

In order to gain a more transparent understanding of the
double-exchange interaction we investigate first 
the case of two sites and one conduction electron, 
as was done originally by Anderson and Hasegawa, and  
de Gennes.\cite{detheory} 
In the case of ferromagnetic coupling ($J < 0$) the 
ground state energy is 
$E_{0, \: J < 0} = - \vert J \vert / 4 - t$ with wave function
$\vert \psi_{0} \rangle_{J < 0} \equiv 
\vert \psi_{\rm DE} \rangle_{J < 0, \: z} = 
\vert \Uparrow_{z} \uparrow_{z}, \:  \Uparrow_{z} 0 \rangle + 
\vert \Uparrow_{z} 0, \: \Uparrow_{z} \uparrow_{z} \rangle$,
where $\Uparrow_{z}$ and $\uparrow_{z}$ refers to the $z$ 
component of the impurity and conduction electron spins, 
respectively. As can be seen, ferromagnetism arises here 
via an Ising type 
coupling, which allows for description of the ground state 
within a simple semiclassical approximation.\cite{detheory} 

For $J > 0$ the situation is completely changed due to the 
singlet formation of localized and conduction electron spins. 
Due to this Kondo singlet formation, which is absent for
the $J < 0$ case, double-exchange is
often ignored in the discussions of the $J > 0$ KLM. 
This case has usually been discussed in terms of the
competition between Kondo singlet formation and the 
RKKY interaction. For a half filled band, this point
of view appears to be sufficient. However for a 
partially filled conduction band an overwhelming 
amount of numerical data \cite{numerics} proves
that ferromagnetism appears for stronger coupling. 

This cannot be explained in terms of 
RKKY, which operates at weak-coupling, nor in 
terms of Kondo singlets, since they are non-magnetic. The missing 
element is double-exchange ordering \cite{graeme} due to an 
excess of localized spins over conduction electrons.
Double-exchange requires only that the number of conduction
electrons is less than the number of localized spins. It 
operates in any dimension, for any sign of 
the coupling $J$, and for any magnitude $S_{j}^{2}$ 
of the localized spins.\cite{detheory}. 
The double-exchange
interaction is specific to the Kondo lattice, and 
is absent in single- or dilute-impurity 
systems in which the situation is reversed, and the 
electrons greatly outnumber the localized spins. 

Double-exchange is conceptually a very simple 
interaction: each electron has on average 
more than one localized spin to screen, and 
consequently hops between several adjacent spins 
gaining screening energy at each site, together with 
a gain in kinetic energy. Since hopping is 
energetically most favourable for electrons which preserve 
their spin as they hop (called coherent hopping), 
this tends to align the 
underlying localized spins.\cite{detheory}. 

For two sites, 
this causes a mixing of the total spin and an enhancement 
of the Hilbert space, where now 16 elements have to be 
considered. The ground state energy is given by 
$E_{0, \: J > 0} = - J / 4 - 
\sqrt{J^{2} + 2 J t + 4 t^{2}} / 2$ with wave functions
$\vert \psi_{0} \rangle_{J > 0} \propto \vert \psi_{\rm KS} 
\rangle_{z} + [ 1 / (J/4 - E_{0, \: J > 0}) ] \: 
\{ \vert \Uparrow_{z} \downarrow_{z}, \: \Uparrow_{z} 0 \rangle 
+ \vert \Uparrow_{z} 0, \: \Uparrow_{z} \downarrow_{z} \rangle
- \vert \Uparrow_{z} \uparrow_{z}, \: \Downarrow_{z} 0 \rangle 
- \vert \Downarrow_{z} 0, \: \Uparrow_{z} \uparrow_{z} \rangle \}$,
where the Kondo singlet $\vert \psi_{\rm KS} \rangle_{z}$ states are
$ \vert \Uparrow_{z} \downarrow_{z}, \: \Uparrow_{z}, 0 \rangle 
- \vert \Downarrow_{z} \uparrow_{z}, \: \Uparrow_{z} 0 \rangle
+ \vert \Uparrow_{z} 0, \: \Uparrow_{z} \downarrow_{z} \rangle
- \vert \Uparrow_{z} 0, \: \Downarrow_{z} \uparrow_{z} \rangle$. 
$\vert \psi_{0} \rangle_{J > 0}$ involves six basis elements 
(the degeneracy is partially lifted by conduction electron hopping) 
and hence falls outside the four dimensional space needed to 
establish double-exchange for $J < 0$. 

In order to invoke double-exchange as well, all 
three spin directions, $x$, $y$, and $z$, have to be considered: 
$\vert \psi_{0} \rangle_{J > 0} \propto 
[1 - 1 / (J/4 - E_{0, \: J > 0}) ] \: \vert \psi_{\rm KS} \rangle_{z} 
+ [ 1 / (J/4 - E_{0, \: J > 0}) ] \: 
\{ \vert \psi_{\rm DE} \rangle_{J > 0, \: x} 
+ \vert \psi_{\rm DE} \rangle_{J > 0, \: y}
+ \vert \psi_{\rm DE} \rangle_{J > 0, \: z} \}$, where
$\vert \psi_{\rm DE} \rangle_{J > 0, \: \alpha = x \: {\rm or} \: y} = 
\{ \vert \Uparrow_{\alpha} \downarrow_{\alpha}, \:  \Uparrow_{\alpha} 0 \rangle 
+ \vert \Uparrow_{\alpha} 0, \:  \Uparrow_{\alpha} \downarrow_{\alpha} \rangle
+ \vert \Downarrow_{\alpha} \uparrow_{\alpha}, \: \Downarrow_{\alpha} 0 \rangle
+ \vert \Downarrow_{\alpha} 0, \: \Downarrow_{\alpha} \uparrow_{\alpha} \rangle
\} / {\sqrt{2}}$ and 
$\vert \psi_{\rm DE} \rangle_{J > 0, \: \alpha = z} = 
  \vert \Uparrow_{z} \downarrow_{z}, \: \Uparrow_{z} 0 \rangle
+ \vert \Uparrow_{z} 0, \: \Uparrow_{z} \downarrow_{z} \rangle$. 
In spite of this extra complication, it is apparent from the above 
that also for $J > 0$, ferromagnetism appears. 

We show in Section V that coherent conduction 
electron hopping over a characteristic length $\lambda$ 
may be incorporated into a bosonization description which 
keeps the electrons finitely delocalized. At lengths 
beyond $\lambda$, the electrons 
are described by collective density fluctuations,  
common to one dimensional Fermi systems. 
The electrons remain finitely delocalized  
over shorter lengths, and describe coherent hopping over several 
adjacent sites. This tends to align  
the underlying localized spins at stronger coupling. 
$\lambda$ measures the effective 
range of the double-exchange interaction, and hence 
is proportional to the width of the magnetic polarons. 

\section{The model}

The KLM considers the coupling between a half-filled narrow 
band (localized $d$ or $f$) and conduction electrons. Even 
though studied intensively for the last two decades, 
the understanding of the KLM remains incomplete. Only in 
one dimension have numerical simulations \cite{numerics} and 
bosonization techniques \cite{boso,graeme} been 
carried through to admit predictions about the phase 
diagram of the KLM. No investigations exist for the case 
where the KLM is extended to account for contributions 
stemming from the phonons, which is of special relevance 
to CMR materials. In particular, the small doping regime of these 
systems, which are ferromagnetic at low temperatures, 
seems to be appropriate for modelling within the KLM 
argumented by interactions with the lattice. In the 
following we present bosonized solutions of the KLM 
where on-site Coulomb and specific phonon contributions are 
explicitly included. This ``extended'' KLM model allows 
spin-and magnetoelastic-polaron formation, which we believe are 
of major importance in understanding these complex materials. 

The Hamiltonian of the KLM in the presence of on-site Coulomb 
interaction reads: 
\begin{eqnarray}
H_{\rm KLM+U} \: &=& \: -t \sum_{j, \sigma} (c^{\dagger}_{j, \sigma} 
c^{}_{j+1, \sigma} + {\rm{h.c.}}) 
\nonumber \\
&+& \: J \sum_{j} {\bf S}_{{\rm d}, j} 
{\bf \cdot} {\bf S}_{{\rm c}, j} + U \sum_{j} n_{j, \uparrow} 
n_{i, \downarrow} \; , 
\label{one}
\end{eqnarray}
where $t > 0$ is the conduction electron hopping integral,
${\bf S}_{{\rm d}, j} = \frac{1}{2} \sum_{\sigma^{},\sigma^{\prime}}
c^{\dagger}_{{\rm d}, j, \sigma^{}} {\vektor{\sigma}}_{\sigma^{}, 
\sigma^{\prime}} c^{}_{{\rm d}, j, \sigma^{\prime}}$, 
${\bf S}_{c, j} = \frac{1}{2} \sum_{\sigma^{},\sigma^{\prime}} 
c^{\dagger}_{j, \sigma^{}} {\vektor{\sigma}}_{\sigma^{},
\sigma^{\prime}} c^{}_{j, \sigma^{\prime}}$ and 
${\vektor{\sigma}}$ are the Pauli spin matrices.  Fermi 
operators $c^{}_{j, \sigma}, c^{\dagger}_{j, \sigma}$ with 
subscript $d$ refer to localized $d$-spins, while those not 
indexed refer to the conduction electrons. The on-site Coulomb 
repulsion is given by the Hubbard term proportional to $U$. 
In the CMR materials the localized states are represented by 
the threefold degenerate Mn $t^{3}_{2g}$ $d$-electrons with total 
spin 3/2. However, for reasons of transparency, the localized spin 
is approximated here by spin 1/2. The properties of the model 
are qualitatively independent of the magnitude of the localized spins,
i.e., the basic features of the phase diagram for the three 
dimensional case appear in one dimensions as well \cite{yuno1,Martin-Mayor}.

In the following the Kondo coupling $J$ is measured in units 
of the hopping $t$ and both cases, antiferromagnetic ($J > 0$) 
and ferromagnetic ($J < 0$) couplings, will be considered. The 
conduction band filling is given by $n = N_{c}/N < 1$, where $N$ 
is the number of lattice sites and $N_{c}$ is the number of
conduction electrons. In order to understand the broad 
properties of CMR materials, we also allow 
for the number of impurity spins,  $N_{d}$, to vary, 
in such a way that $N_{d} / N < 1$. In this way, doped or dilute
Kondo lattice systems can also be studied. It will be shown
that a dilute Kondo lattice system is dominated by short range 
antiferromagnetic correlations. 

In one dimension, the electron-phonon coupling can either be of  
inter-site (Su-Schrieffer-Heeger (SSH) \cite{ssh}) or on-site
(Holstein \cite{holstein}) character. 
The models we study assumes a 
dispersionless phonon mode with frequency $\omega$.
The neglect of the dispersion of bare phonons is not essential since 
it is absent in the Holstein model and the acoustic phonons are decoupled
from the low energy electronic spectrum in the continuum limit of the 
SSH model.\cite{hirsch}
In this approximation the Holstein 
coupling to dispersionless phonons has the following form: 
\begin{equation}
H_{\rm Holstein} = \sum_{i} \Big( \beta q^{}_{i} n^{}_{ci} + 
\frac{K}{2} q^{2}_{i} + 
\frac{1}{2 M} p^{2}_{i} \Big) \; , 
\label{ph-egy}
\end{equation}
with the conduction electron density $n^{}_{c i}$ at site $i$, the 
lattice displacement $q^{}_{i}$, its conjugate momentum $p^{}_{i}$, 
the electron-phonon coupling strength $\beta$, the spring constant 
$K$ and the ionic mass $M$. Within the SSH model the coupling to 
phonons is modified to 
\begin{equation}
H_{\rm SSH} = 
\sum_{i} \Big[ \sum_{\sigma} \alpha_{\sigma} (q^{}_{i + a} - q^{}_{i}) 
(c^{\dagger}_{c i \sigma} c^{}_{c i + a \sigma} + 
c^{\dagger}_{c i + a \sigma} c^{}_{c i \sigma}) 
+ \frac{K}{2} (q^{}_{i + a} - q^{}_{i})^2 
+ \frac{1}{2 M} p^{2}_{i} \Big] \; , 
\label{ph-ot}
\end{equation}
where $\alpha_{\sigma}$ denotes the electron-phonon coupling strength. 
Thus, the starting hamiltonian is 
\begin{equation}
H \: = \: H_{\rm KLM+U} \: + \: H_{\rm Holstein} \: + \: H_{\rm SSH} \, . 
\label{ph-new-egy}
\end{equation}

These phononic contributions may not describe the full 
complexity of the phononic couplings observed in real materials 
because of the phase space constraint of any one dimensional 
calculation. However, the results capture the essence of the Kondo
lattice coexisting with phonons, and being an exact solution, it 
represents a vital source of information due to the lack of 
comparable solutions applicable to colossal magnetoresistance 
materials. 

\section{The effective hamiltonian} 

A large class of one dimensional many-electron systems may 
be described using bosonization techniques 
\cite{haldane}: The electron fields 
may be represented in terms of collective 
density operators which satisfy  
bosonic commutation relations. Bosonic  
representations provide a non-perturbative
description which, in general, by far easier to evaluate than
a formulation in terms of fermionic operators. 

The underlying bosonization scheme follows standard 
procedures \cite{haldane} by first decomposing the 
on-site operators into Dirac fields, 
$c^{}_{x, \sigma} \: = \: 
\sum_{\tau} e^{i k_{F} x} \Psi_{\tau, \sigma} (x)$, 
where $k_{F} = \pi n /2$, 
with spinor components $\tau = \pm$ 
(+/- being the right/left movers) and $k_{F} = \pi n / 2$. 
Next we bosonize the Dirac fields with $\Psi_{\tau, \sigma} 
= \exp (i \Phi_{\tau, \sigma}) 
/ \sqrt{2 \pi \lambda}$, where $1 / \lambda$ is the ultraviolet 
cutoff. For the scalar Bose fields, $\Phi_{\tau, \sigma} (x)$,  
and their conjugate momenta, $\Pi_{\tau, \sigma} (x)$, 
$\Phi_{\tau, \sigma} (x) = \int^{x}_{-\infty} d x^{\prime} 
\Pi_{\tau, \sigma} (x^{\prime})$, are used in standard 
Mandelstam representation by means of which a momentum 
cutoff via the Fourier transform is introduced
$\Lambda (k) = \exp ( - \lambda \vert k \vert / 2 )$. 
If the distance between the impurity spins is larger than
$\lambda$, the electrons will behave as collective 
density fluctuations.\cite{haldane} Thus, the 
Fermi fields can be represented in terms of density
operators which satisfy Bose commutation relations: 
\begin{equation}
c^{}_{\tau, x, \sigma} \: = \: \exp ( i \tau k_{F} x) 
\exp i \{ \theta_{\rho}(x) + \tau \phi_{\rho}(x) 
+ \sigma [ \theta_{\sigma}(x) + \tau \phi_{\sigma}(x)] \} / 2 \; ,
\label{diag}
\end{equation} 
where the Bose fields for $\nu = \rho, \sigma$ are defined by
\begin{equation}
\psi_{\nu}(x)  \: = \: i (\pi / N) \sum_{k \ne 0} e^{i k x}
[ \nu_{+}(k) \pm \nu_{-}(k) ] \Lambda(k) / k \; , 
\label{boso}
\end{equation}
with $+$ 
corresponding to the number fields $\psi_{\nu} = \phi_{\nu}$ 
and $-$ to the current fields $\psi_{\nu} = \theta_{\nu}$.
The charge (holon) and spin (spinon) number fluctuations are 
defined as $\rho_{\tau}(k) = \sum_{\sigma} \rho_{\tau, \sigma}(k)$, 
and $\sigma_{\tau}(k) = \sum_{\sigma} \sigma \rho_{\tau, \sigma}(k)$. 
All rapidly oscillating terms originating from e.g. backscattering 
and umklapp processes are neglected, since they contribute only at 
exactly half filling. The localized $d$ electrons can neither be 
bosonized nor Jordan-Wigner transformed since no direct interaction 
exists between them. 

Substituting these representations into  
Eq.\ (\ref{one}) gives the bosonized KLM hamiltonian
\begin{eqnarray}
H_{\rm KLM+U} &=& \frac{1}{4\pi} \sum_{j,\nu} v_{\nu} 
\left\{ \Pi_{\nu}^{2}(j) + [\partial_{x}\phi_{\nu}(j)]^{2}
\right\} + 
\frac{J}{2 \pi}\sum_{j} [\partial_{x}\phi_{\sigma}(j)] 
S_{{\rm d}, j}^{z}
\nonumber \\
&+& \: \frac{J}{4 \pi \lambda} \sum_{j} \left\{
\cos [\phi_{\sigma}(j)] + \cos[2k_{F}j + \phi_{\rho}(j)] 
\right\} \left(e^{-i \theta_{\sigma}(j)} S_{{\rm d}, j}^{+} 
+ {\rm h.c.} \right)
\nonumber \\
&-& \: \frac{J}{4 \pi \lambda} \sum_{j} \: \sin[\phi_{\sigma}(j)]
\sin[2k_{F}j + \phi_{\rho}(j)] S_{{\rm d}, j}^{z} \; .  
\label{bklm}
\end{eqnarray}
The charge and spin velocities are given below. 

Considering the phononic contributions, 
in standard bosonization language, Eq.\ \ref{ph-egy} 
simplifies to $H^{\rm ph} + H^{\rm el-ph}_{1} + H^{\rm el-ph}_{2}$,
where 
\begin{eqnarray}
H^{\rm ph} \: &=& \: \frac{1}{2 N} \sum_{p} \left[ \Pi^{2}_{0} (p) + 
\omega^{2}_{0} \Phi^{2}_{0} (p) \right]
\nonumber \\
&+& \: \frac{1}{2} \int dx \left[ \Pi^{2}_{\pi} (x) + 
\omega^{2}_{\pi} \Phi^{2}_{\pi} (x) \right] \, ,
\label{ph-ketto}
\end{eqnarray}
and $\omega_{0} = \omega_{\pi} = \sqrt{ K / M}$ are their 
respective phonon frequencies. 
The electron-phonon forward scattering term is simply
\begin{equation}
H^{\rm el-ph}_{2} \: = \: \gamma_{2} \frac{{\sqrt2}}{N} \sum_{p} 
[ \rho_{+} (-p) + \rho_{-} (-p) ] \Phi_{0} (p) \; . 
\label{ph-harom}
\end{equation}
On the other hand, 
the rapidly oscillating phonon-assisted backward scattering
term will acquire an extra factor $\exp [ \pm i \pi \delta n x]  
\equiv \exp [ \pm i (2 k_F - \pi) x ]$, in the form:  
\begin{equation}
H^{\rm el-ph}_{1} \: = \: \gamma_{1} \sum_{\nu = \pm, \sigma} \int dx 
[ \Psi^{\dagger}_{\nu, \sigma} \Psi^{}_{-\nu, \sigma} 
e^{i \pi \nu \delta n x} ] \Phi_{\pi} (x) \; , 
\label{ph-negy}
\end{equation}
with $\gamma_{1} = \gamma_{2} = \beta / \sqrt{M}$, where we used 
the same subscripts for backward and forward scattering  as in 
g-ology.\cite{haldane}

In the continuum limit, the SSH
term in contrast to the Holstein coupling, 
gives only two terms $H^{\rm ph} + H^{\rm el-ph}_{- 1}$ - 
the standard phononic component and a rapidly oscillating  back 
scattering term. $H^{\rm ph}$ 
is given in Eq.\ \ref{ph-ketto}, while the back scattering term 
$H^{\rm el-ph}_{- 1}$ differs from Eq.\ \ref{ph-negy} only
in a form factor:
\begin{equation}
H^{\rm el-ph}_{- 1} \: = \: \gamma_{- 1} \sum_{\nu = \pm, \sigma} 
\int dx [ i \nu \Psi^{\dagger}_{\nu, \sigma} \Psi^{}_{-\nu, \sigma} 
e^{i \pi \nu \delta n x} ] \Phi_{\pi} (x) \; ,
\label{ph-negy-new}
\end{equation}
with $\gamma_{- 1} = 4 \alpha_{\sigma} / \sqrt{M}$. 
However, the fact that the forward scattering term is missing 
means that the SSH coupling will not give any contribution to
the effective hamiltonian away from half filling. 

Thus, the transformed hamiltonian of Eq.\ (\ref{ph-new-egy}) is: 
\begin{eqnarray}
H  \: &=& \: \frac{1}{4\pi} \sum_{j,\nu} \: v_{\nu} \: 
\left\{ \Pi_{\nu}^{2}(j) + [\partial_{x}\phi_{\nu}(j)]^{2}
\right\} \: + \: 
\frac{1}{2 N} \sum_{p} \left[ \Pi^{2}_{0} (p) + 
\omega^{2}_{0} \Phi^{2}_{0} (p) \right] 
\nonumber \\
&+& \: \gamma_{2} \frac{{\sqrt2}}{N} \sum_{p} 
[ \rho_{+} (-p) + \rho_{-} (-p) ] \Phi_{0} (p) 
\: + \: \frac{1}{2} \int dx \left[ \Pi^{2}_{\pi} (x) + 
\omega^{2}_{\pi} \Phi^{2}_{\pi} (x) \right] 
\nonumber \\
&+& \: \frac{J}{4 \pi \lambda} \sum_{j} \left\{
\cos [\phi_{\sigma}(j)] + \cos[2k_{F}j + \phi_{\rho}(j)] 
\right\} \left(e^{-i \theta_{\sigma}(j)} S_{{\rm d}, j}^{+} 
+ {\rm h.c.} \right)
\nonumber \\
&-& \: \frac{J}{4 \pi \lambda} \sum_{j} \: \sin[\phi_{\sigma}(j)]
\sin[2k_{F}j + \phi_{\rho}(j)] S_{{\rm d}, j}^{z} 
\: + \: 
\frac{J}{2 \pi}\sum_{j} [\partial_{x}\phi_{\sigma}(j)] 
S_{{\rm d}, j}^{z} \; .
\label{two}
\end{eqnarray} 
If holes are present in the array of $d$-spins, all terms 
proportional to $S$ are zero. The charge and spin velocities are
\begin{equation}
v_{\rho} = v_{F} \sqrt{1 + U /\pi v_{F} - \beta^2 / \pi K v_{F}} \,\, ,
\quad \quad 
v_{\sigma} = v_{F} \sqrt{1 - U/\pi v_{F} + \beta^2 / \pi K v_{F}} \,\, ,      
\label{velocities} 
\end{equation}
where the Fermi velocity is $v_{F} = 2\sin(\pi n/2)$ in units of $t$.

It is important to note that a renormalization of the 
spinon-holon velocities appears here due to the Hubbard 
and phonon terms which act oppositely on the corresponding 
velocities. While the Hubbard term leads to a localization 
of the spinons and an increased hopping of the holons, thus 
supporting a magnetic ground state, the phonons delocalize 
the spins, but localize the charges and act destructively 
on the magnetic properties. It is worth mentioning that the 
Hubbard term alone already suffices to establish two time 
scales for the holon-spinon dynamics. But an important 
renormalization of the critical properties of the system 
is achieved through the variable phonon coupling, which, 
as will be shown below, establishes the existence of a 
Griffiths phase. The competition between the Hubbard and 
the phonon term obviously vanishes for $U = \beta^2 / K$.  

In the following, effects arising from the localized spin 
$d$ impurities, double-exchange, the phonons and Hubbard 
interactions will be discussed in more detail. The localized 
spin $d$ impurities act via double-exchange on the hopping 
electrons so as to preserve their spin when moving through the 
lattice in order to screen the localized spins which are in 
excess of the conduction electrons, i.e. $N \le N_{d} > N_{c}$. 
This, in turn, leads to a tendency to align the localized spins 
and results in an additional screening energy for the 
conduction electrons. 

\section{Localized spin ordering}

In order to determine rigorously the phase diagram and 
investigate the ordering of the local 
spins due to the formation of polarons, we firstly apply
a unitary transformation. This is the simplest method
for determining the ordering of the localized spins induced 
by the conduction electrons and it has been used successfully
elsewhere \cite{boso,graeme}. This is achieved by choosing
a basis of states for the unitary transformation in which 
competing effects become more transparent, i.e. a
transformation which changes to a basis of states in which 
the conduction electron spin degrees of freedom are directly 
coupled to the localized spins. Correspondingly, we choose the operator
${{\hat{\rm S}}} = i ( J / 2 \pi ) {\sqrt{v_{F} / v^{3}_{\sigma}}} 
\sum_{j} \theta_{\sigma}(j) \: S^{z}_{{\rm d}, j}$, which 
is applied to Eq.\ (\ref{two}) up to infinite
order, thus avoiding truncation errors. 
 
Secondly, we explicitly take into account the Luttinger 
liquid character of the Bose fields, i.e., use their
non-interacting expectation values such that the 
effective Hamiltonian for the local spins is derived as:
\begin{eqnarray}
H_{{\rm eff}} \: &=& \: - {\frac{J^2 v^{2}_{\sigma}}{4 \pi^2 v_{F}}}
\sum_{j, j^{\prime}} \: \int^{\infty}_{0} dk 
\cos [ k (j - j^{\prime})] \Lambda^{2} (k) \: 
S^{z}_{{\rm d}, j} S^{z}_{{\rm d}, j^{\prime}}
\nonumber \\
&+& \: {\frac{J}{2 \pi \lambda}} \sum_{j}
\{ \cos[K(j)] + \cos[2 k_{F} j] \} S^{x}_{{\rm d}, j}
\nonumber \\
&-& \: {\frac{J}{2 \pi \lambda}} \sum_{j}
\sin[K(j)] \sin[2 k_{F} j] S^{z}_{{\rm d}, j} \; .
\label{heff}
\end{eqnarray}
Here $K(j)$ arises from the unitary transformation and counts all 
the $S^{z}_{{\rm d}, j}$'s to the right of the site $j$ 
and subtracts all those to the left of $j$: $K(j) = 
(J / 2 v_{F}) \sum_{l = 1}^{\infty} ( S^{z}_{{\rm d}, j + l} - 
S^{z}_{{\rm d}, j - l} )$. This term gives the crucial 
difference between the Kondo lattice and dilute Kondo lattice, 
as will be explained in the following. 

\begin{figure} 
\includegraphics[width=5in]{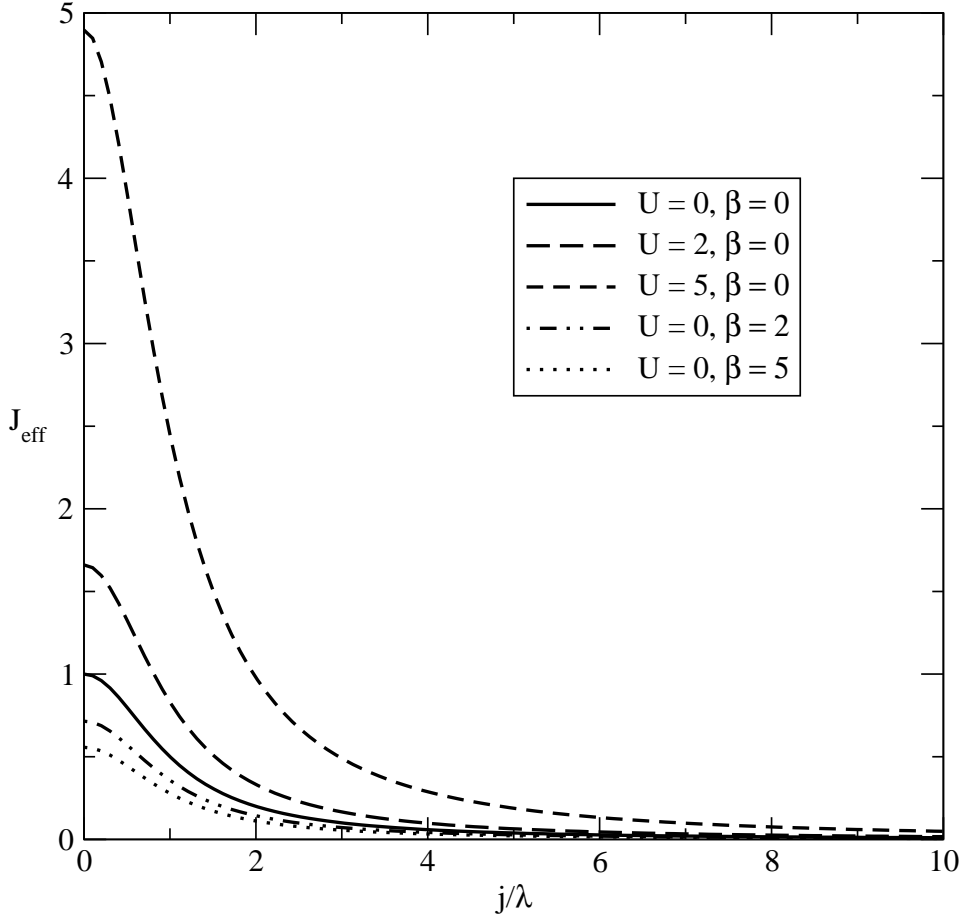} 
\caption{\label{fig:Fig1} The range of the ferromagnetic interaction
Eq.\ (\ref{ferro-longrange}) for different values of $U$ and $\beta$. 
$J_{\rm eff}$ is the interaction strength in units of 
$J^2 v^2_{\sigma} / 4 \pi^2 v_{F}$}. 
\end{figure}

The most important term in Eq.\ (\ref{heff}) is the first one, 
\begin{equation}
J_{\rm eff} (j - j^{\prime}) = {\frac{J^2 v^{2}_{\sigma}}{4 \pi^2 v_{F}}}
\: \int^{\infty}_{0} dk 
\cos [ k (j - j^{\prime})] \: \Lambda^{2} (k) \; .
\label{ferro-longrange}
\end{equation}
This term shows that a double-exchange ferromagnetic interaction
appears for both $J > 0$ and $J < 0$ coupling and even in the 
dilute Kondo lattice model. This coupling is non-negligible 
for $N_{d} > N_{c}$ and $j - j^{\prime} \le \lambda$ and its 
strength decreases with increasing distance between impurity spins. 
The value of $J_{\rm eff}$ is plotted for different values of $U$ 
and $\beta$ in Fig.\ \ref{fig:Fig1}. 

Fig.\ \ref{fig:Fig1} shows clearly that the Hubbard interaction 
enhances double-exchange and consequently $J_{\rm eff}$. As 
noted earlier, see the discussion following Eq.\ (\ref{velocities}), 
this is due to the fact that the Hubbard term leads to a localization
of spinons and an increase in the hopping of holons, hence increasing 
the double-exchange. On the other hand, the electron-phonon interaction 
counteracts this effect by localizing the holons and thus decreasing 
$J_{\rm eff}$. 

\begin{figure} 
\includegraphics[width=5in]{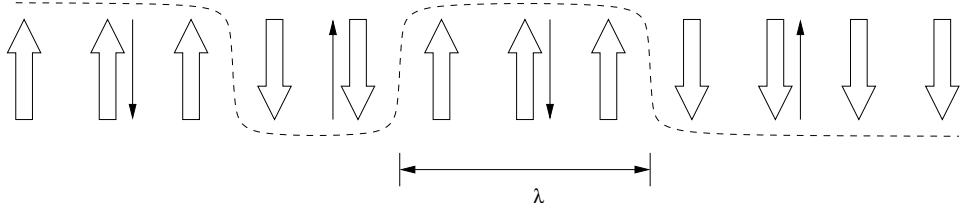} 
\caption{\label{fig:Fig2} A snapshot of the magnetic polarons. 
$\Uparrow$ ($\Downarrow$) and $\uparrow$ ($\downarrow$) refers 
to the $z$ component of the impurity and conduction electrons spins. 
The dashed curve represents the spin domain walls (kink-antikink pairs). 
The electrons will oscillate between these domains walls
(dotted circle) and hence couple the impurity spins through 
double-exchange (for details see text). $\lambda$ is the average 
length of the polarons, i.e., the effective  range of the 
double-exchange.}. 
\end{figure}

Fig.\ \ref{fig:Fig1} also shows that the length $\lambda$ 
characterizes the effective range of the double-exchange interaction. 
This interaction originates from the bosonization of the  
conduction electron band in the following way: at wavelengths 
larger than $\lambda$, the  
electrons combine to form collective density fluctuations, which  
involve large numbers of electrons satisfying  
bosonic commutation relations. This is the standard  
behavior of one dimensional many-electron systems for weak 
interactions \cite{haldane}. At wavelengths smaller than $\lambda$, 
the density fluctuations are not collective, and loose the 
bosonic character. Since bosonization only applies to 
fluctuations beyond $\lambda$, 
the bosonization description is equivalent to keeping the electrons 
finitely delocalized within the range of $\lambda$, with the electrons 
preserving their spin over this range. Eq.\ (\ref{ferro-longrange}) 
describe the ordering induced on the localized spins by the 
finitely delocalized electrons. Thus, $\lambda$ corresponds to the 
effective delocalization length related to the spatial extent 
of the polarons (polaron radius), i.e., the effective range of 
double-exchange, as shown in Fig.\ \ref{fig:Fig2}. 

Double-exchange becomes inefficient if the distance between the impurity 
spins is larger than $\lambda$. In general $\lambda$ has a complex
behavior, depending on $J$, $U$, $\beta$ and $N_{c}$. For large
distances and low density, Eq.\ (\ref{ferro-longrange}) gives an 
effective range for the double-exchange which decays exponentially 
with a characteristics scale $\sqrt{2/ J}$. Hence, in the following 
we will use for $\lambda$ its low density (e.g., in the example 
depicted in Fig.\ \ref{fig:Fig2}) value: $\lambda \approx {\sqrt{2 / J}}$. 
And since the interaction Eq.\ (\ref{ferro-longrange}) is short range for
all finite $\lambda$ we approximate it by its nearest-neighbor form: 
\begin{equation}
J_{\rm eff} = {\frac{J^2 v^2_{\sigma}}{2 \pi^2 v_{F}}}
\int^{\infty}_{0} dk \cos k \: \Lambda^{2} (k) \; .
\label{ferro}
\end{equation}

Here we emphasize again that Eq.\ (\ref{ferro}) is valid for both
$J > 0$ and $J < 0$ couplings. Since the $J > 0$ case is usually 
not considered in discussions of the KLM, 
we emphasize the following points: 
(1) the term originates, via bosonization and 
then the unitary transformation, from the kinetic energy term 
$-t \sum_{j, \sigma} (c^{\dagger}_{j, \sigma} 
c^{}_{j+1, \sigma} + {\rm{h.c.}})$ 
and the forward scattering part of 
$(J/2) \sum_{j} (n_{j\uparrow} - n_{j\downarrow}) 
S^{z}_{j}$ ($n_{j\sigma} = c^{\dagger}_{j\sigma}c^{}_{j\sigma}$) in 
the KLM hamiltonian Eq.\ (\ref{one}). (Note that the Bose 
representations for the electrons in these terms are exact.) 
(2) Eq.\ (\ref{ferro}) is 
independent of the sign of $J$, and takes the same form  
for any magnitude $S_{j}^{2}$ of the localized spins. 
(3) Since Eq.\ (\ref{ferro}) is of order $J^{2}$, 
whereas the remaining terms in the transformed hamiltonian 
Eq.\ (\ref{heff}) are of order $J$, the interaction    
Eq.\ (\ref{ferro}) dominates the ordering of the localized 
spins as $J$ increases. All these properties are identical
to those of a double-exchange interaction. This leads us to
identify Eq.\ (\ref{ferro}) as the double-exchange interaction
in the KLM. We also emphasize
that Eq.\ (\ref{ferro}) is ferromagnetic for all 
(possible) choices of the cut-off function 
$\Lambda (k)$. 

Based on property (1) above, a simple characterization 
can be given for the double-exchange valid at low conduction 
band filling. The simple hamiltonian term (1) from above 
satisfy a standard nonlinear Schr{\"o}dinger equation:  
$\partial^{2}_{x} \psi_{\sigma}(x) + (J m_{{\rm el.}} / 2) 
\vert \psi_{\sigma}(x) \vert^{2} \psi_{\sigma}(x)
= 2 m_{{\rm el.}} E \psi_{\sigma}(x)$
($m_{{\rm el.}}$ being the bare electron mass, and $\psi_{\sigma}(x)$
the electronic wave function) 
with soliton solutions $\psi_{\sigma}(x) \propto  e^{i x} 
{\rm sech} ( x \sqrt{J m_{{\rm el.}}  / 4} )$; see 
Ref.\ \onlinecite{Makhankov}. 

These soliton solutions correspond to spin domain walls of 
finite size (kink-antikink pairs) and lead to a gain 
in electronic energy of $- \sigma$ for antiferromagnetic coupling, 
and of $+ \sigma$  for the ferromagnetic case, 
as shown in Fig.\ \ref{fig:Fig2}
Physically the solutions resemble the dressing of the electron 
by a finite range of parallel (antiparallel) local spins and 
consequently can be identified with polaronic type objects. 
From the previous considerations 
it can also be concluded that, when including the interactions 
with the phonons, the tendency towards charge localization is 
enhanced and increases this polaronic effect. Since the lattice 
also experiences a renormalization due to the coupling to 
the electronic degrees of freedom, substantial ionic 
displacement patterns will develop and the formation 
of magnetoelastic polarons takes place. Similar results 
are obtained by decoupling electronic and phononic degrees 
of freedom through a homogeneous Lang-Firsov transformation, 
where the localization stems from band narrowing. 
In accordance with previous results, 
the polaron radius is characterized by 
a length scale proportional to ${\sqrt{2 / J}}$ . 
This new length scale differs from the free 
conduction electrons mean free path 
and gives rise to competing time scales: slow motion of 
the polaronic carriers and fast motion of the free electrons,  
thus providing dynamics of two types of particles and a 
close analogy to a two fluid scenario.\cite{curro} Since 
the polarons are in general randomly 
distributed within the local spin array, these states 
can be viewed as intrinsic inhomogeneities involving 
spin fluctuations and short-range spin correlations. 
In addition, these new slow dynamics will exhibit a peak 
in the spin structure factor at $2 k_F - \pi$ instead of 
the simple $2 k_F$ RKKY signal. A similar observation 
has also been made \cite{numerics} using numerical approaches. 

\section{The phase diagram}

For the Kondo lattice model, $K(j) \approx 0$ as the number of 
localized spins to the left and the right of a given 
site $j$ are the same. If, however, we have a small 
concentration of holes in the array of localized 
spins, then - opposite to the previous case - $K(j)$  
is non-vanishing since the hole spins are no longer 
necessarily equally 
distributed to the left and the right of a given site. 
This yields $K(j) \approx (-1)^{j} (J / 2 v_{F})$,
which gives rise to a staggered field and antiferromagnetic 
ordering.\cite{dilute}

Since our main interest here is to explore the occurrence of 
ferromagnetism in the presence of the Hubbard and phonon terms, 
we take $K(j)$ vanishingly small and 
focus on the transition between the paramagnetic and the 
ferromagnetic state controlled by the first two terms of Eq.\ (\ref{heff}). 
Hence $H_{\rm eff}$ reduces to a quantum transverse-field Ising chain. 
This model without backscattering is known \cite{pfeuty}
to undergo a quantum phase transition from a ferromagnetic
to a paramagnetic phase. In the case of Eq.\ (\ref{heff}), where
a backscattering is also present, a similar phase transition occurs
as the coupling $J$ is decreased: 
the conduction electrons become less strongly bound to the localized 
spins, and tend to extend over spatial ranges beyond 
the effective range $\lambda$ of double-exchange ordering. 
Double-exchange becomes less effective, i.e., the magnetic polarons
are loosely bound and regions of ordered localized spins begin to 
interfere as the conduction electrons become extended. 
The interference leads to spin-flip processes given 
by the transverse field in $H_{\rm eff}$. 

The transverse-field, $h(j) = (J / 2 \pi \lambda) 
[1 + \cos(2 k_{F} j)]$, in $H_{\rm eff}$, 
includes two low-energy spin-flip processes by means of
which the conduction electrons disorder the localized spins. 
One spin-flip process is backscattering, and is accompanied 
by a momentum transfer of $2k_{F}$ from the conduction 
electrons to the localized spins. Since the chain of localized 
spins will tend to order so as to reflect this transfer, the 
transverse-field corresponding to backscattering spin-flips 
is sinusoidally modulated by $2k_{F}$. The other low-energy 
spin-flip process in $H_{\rm eff}$ is forward scattering. 
This involves zero momentum transfer to the localized spins, 
and the corresponding transverse-field is a constant (i.e.\ 
has modulation zero). 

However, away from half filling the transverse 
field $h(j)$ will have an incommensurate modulation $2k_{F}$ 
with respect to the underlying lattice of localized spins. 
Hence, the conduction band is unable to either totally order 
or totally disorder the lattice as the ferromagnetic to
paramagnetic transition occurs. There remain dilute 
regions of double-exchange ordered localized spins
or magnetic polarons in the paramagnetic phase as 
only a quasi-commensurate fraction of the conduction electrons 
become weakly-bound, and become free to scatter along the chain. 
The remaining ordered regions are sufficiently dilute to prevent
long range correlations, but their existence dominates the 
low-energy properties of the localized spins near the transition.  

These considerations motivated us to treat the transverse field as
a random variable. The effective Hamiltonian can thus be 
replaced by a random transverse field Ising model:\cite{graeme}  
\begin{equation}
H_{{\rm crit}} = - J_{\rm eff} \sum_{j} S^{z}_{{\rm d}, j} 
S^{z}_{{\rm d}, j + \ell} - \sum_{j} h_{j} S^{x}_{{\rm d}, j} \; , 
\label{heff_final}
\end{equation}
where $J_{\rm eff}$ is given in Eq.\ (\ref{ferro}) and 
the ferromagnetic coupling strictly vanishes if 
$\ell > \lambda$. The random fields, $h_{j}$, are generated 
by $(1 + \cos[2 k_{F} j])$ at large distances, where 
$\cos[2 k_{F} j]$ oscillates unsystematically with respect
to the lattice. The large values $\cos[2 k_{F} j] \approx 1$ 
which are responsible for spin flips, are then well separated
and are driven by a cosine distribution analogous to  
spin-glasses.\cite{abrikosov} 

\begin{figure} 
\includegraphics[width=5in]{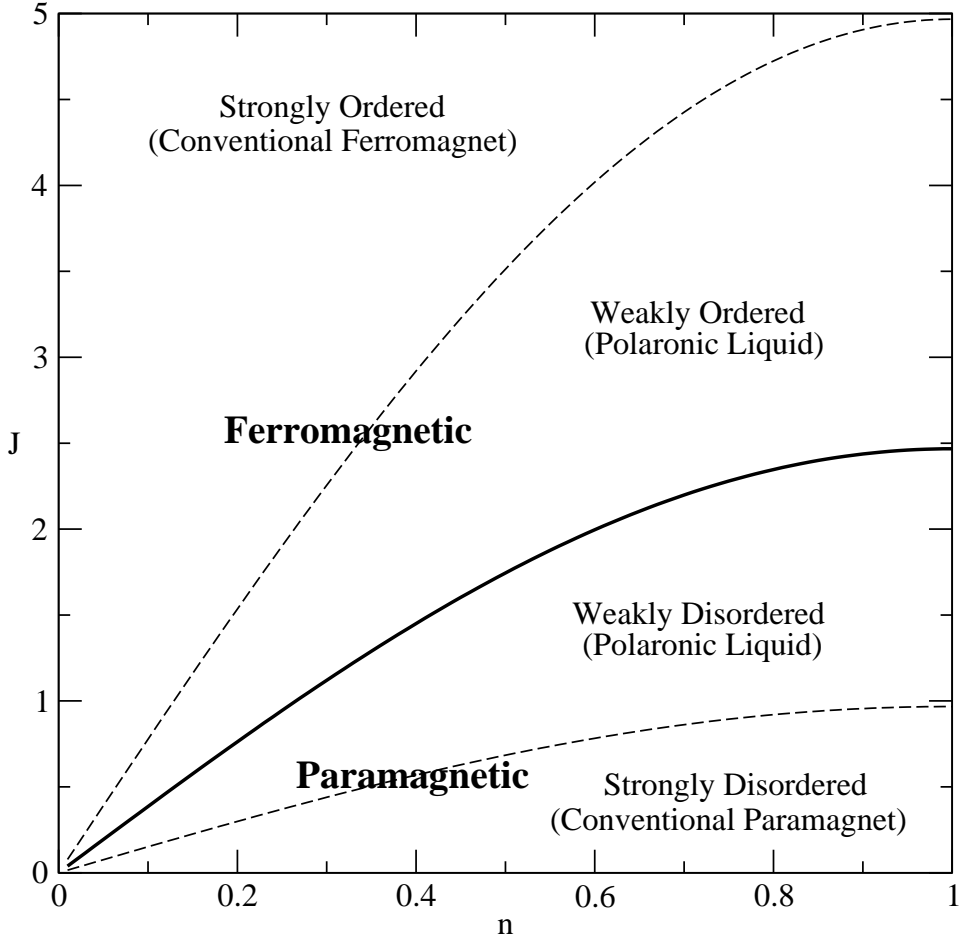} 
\caption{\label{fig:Fig3} The ground-state phase diagram of the 
standard KLM. The solid (critical) line is from Eq.\ (\ref{jcrit}). 
The dashed lines separate strongly ordered (conventional ferromagnet) 
/ disordered (conventional paramagnet) phases from their weak 
Griffiths phase (polaronic liquid) counterparts.} 
\end{figure} 

Using the extensive real space renormalization group results of 
this model by Fisher, Ref.\ (\onlinecite{fisher}), we determine 
the location of the quantum critical line describing
the paramagnetic to ferromagnetic transition at 
\begin{equation}
J_{{\rm crit}} = (\pi^2 / 4) \sin ( \pi n / 2 ) \: 
\{ 1 - U / [2 \pi sin (\pi n / 2)] +  
\beta^2 / [2 \pi K \sin(\pi n / 2)] \}^{1/2} \; .
\label{jcrit}
\end{equation}
For values $J < J_{\rm crit}$ a paramagnetic state exists 
which is dominated by polaronic fluctuations. For $J > J_{\rm crit}$
ferromagnetism appears. The transition between these phases is 
of order-disorder type with variable critical exponent $\delta =
\ln  (J_{\rm crit} / J)$. 

The behavior described by $H_{\rm crit}$
is simply understood in terms of magnetic polarons introduced
previously, see also Fig.\ \ref{fig:Fig2}. Considering for
the moment the standard KLM model, a summary of the known
results \cite{graeme} is presented in Fig.\ \ref{fig:Fig3}. 
Reducing $J$ from intermediate values in the ferromagnetic
phase, the infinite cluster characterizing strong ferromagnetism is
broken up into several large clusters as the quantum
fluctuations $h_{j}$, controlled by the spin-flip interactions,
become stronger. The individual clusters are the spin polarons
depicted in Fig.\ \ref{fig:Fig2}. These magnetic polarons
are weakly ordered in this phase, i.e., form a polaronic
liquid which exists for 
$-0.7 < \delta < 0$ hence with the upper boundary determined 
by $J \approx 2 J_{\rm crit}$, see Fig.\ \ref{fig:Fig3}.
This is not a true transition line, but rather marks
the crossover to a Griffiths phase \cite{griffiths} characterized
by singularities in the free energy over the whole range of $\delta$. 
For small $\delta$ the correlation length is $\xi \sim \delta^{-2}$,
beyond which the system is ordered. The spontaneous
magnetization $M_{0} \propto | \delta |^{\gamma}$,
with $\gamma = (3 - {\sqrt{5}})/2 \approx 0.38$,\cite{fisher} 
while for small applied fields $H$ the magnetization 
$M(H) \propto M_{0}[1+ {\cal{O}}(H^{2|\delta|}\delta \ln H)]$; 
the susceptibility is infinite with a continuously variable exponent. 
The mean correlation function is given by $(\xi/x)^{5/6} e^{-x/\xi}
\exp[-3(\pi x/\xi)^{1/3}]$ for $x \gg \xi$.\cite{fisher} 

\begin{figure} 
\includegraphics[width=5in]{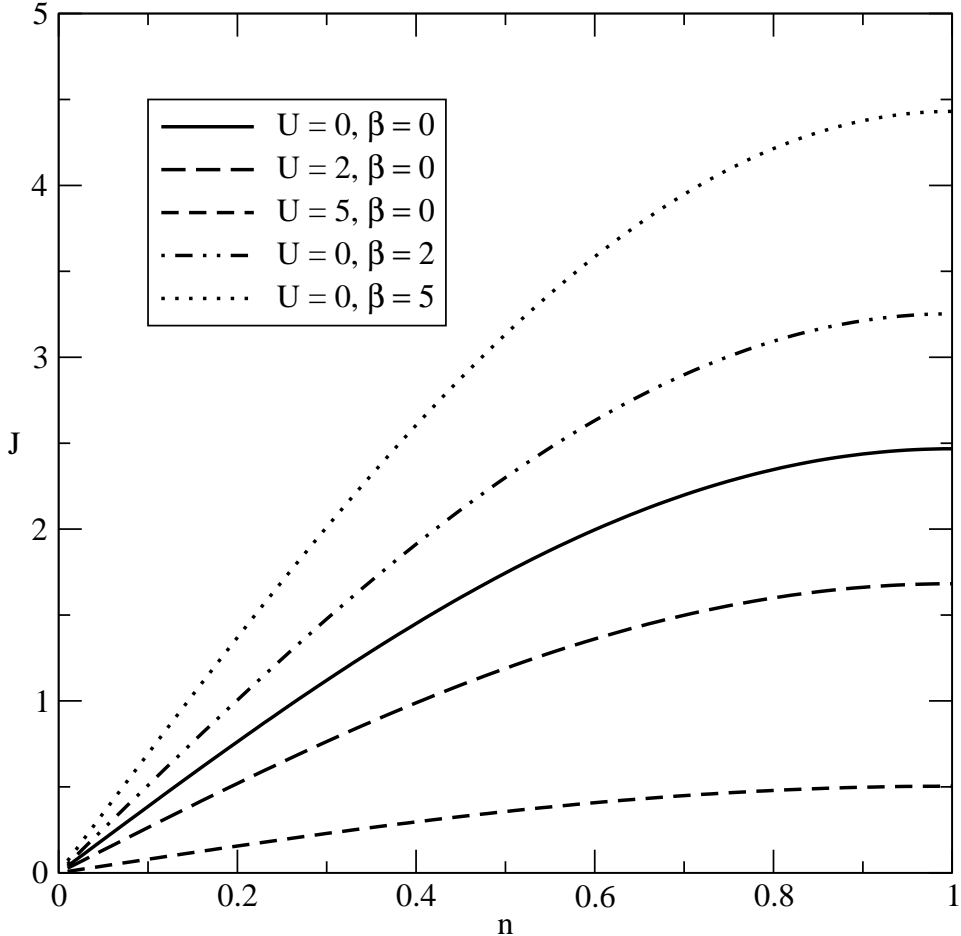} 
\caption{\label{fig:Fig4} The quantum critical line Eq.\ (\ref{jcrit}) 
describing the ferromagnetic transition for different values of $U$ 
and $\beta$. For each of these examples the ferromagnetic phase is 
above the phase transition curve.} 
\end{figure}

Further lowering $J$, we reach the true phase transition
Eq.\ (\ref{jcrit}). The correlation length is infinite, the
magnetization $M(H) \propto |\ln H|^{-\gamma}$ for small $H$, and
the mean correlation function is critical 
$x^{-\gamma}$. The transition curve for different values of
$U$ and $\beta$ is shown in Fig.\ \ref{fig:Fig4}. As argued
previously, the Hubbard interaction makes the ferromagnetic 
phase more robust, as it increases the strength of the 
double-exchange and as such the length of the magnetic polarons. 

\begin{figure} 
\includegraphics[width=5in]{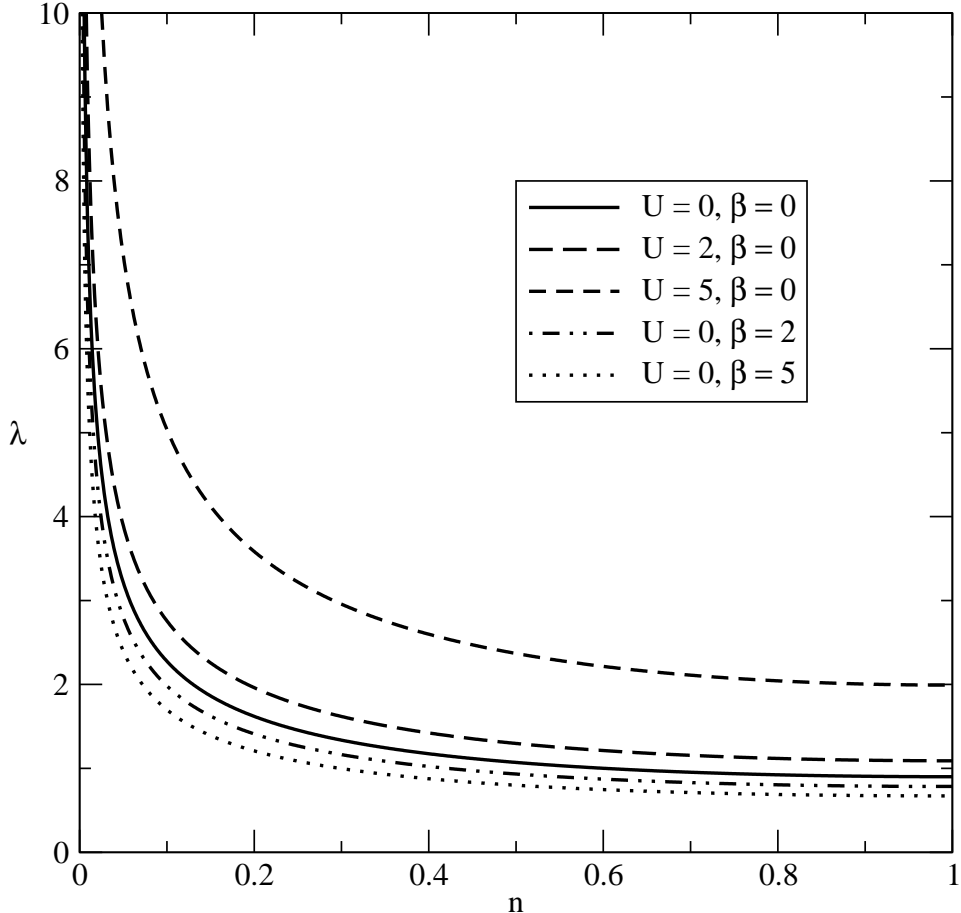} 
\caption{\label{fig:Fig5} The effective range $\lambda$ of 
the double-exchange interaction in units of the lattice 
spacing against filling $n$ on the critical line.} 
\end{figure}

Calculating the effective length of polarons, $\lambda$ on the
transition line, see Fig.\ \ref{fig:Fig5}, supports this finding.
The length of the polarons increases strongly with $U$, hence
making the ferromagnetic phase more difficult to break up. However,
the consequence of this effect is that the Griffiths phase below
the transition will diminish strongly or vanish in most cases.

The effect of phonons is opposite to this. The phonons act destructively
on the magnetic properties. The ferromagnetic phase becomes smaller, 
see Fig.\ \ref{fig:Fig4} and the length of the polarons are slightly
decreased in the presence of the phonons as shown in Fig.\ \ref{fig:Fig5}. 
But, as will be explained later, the boundaries of the lower 
Griffiths phase will remain the same. Thus, contrary the the effect of $U$, 
where the Griffiths phase vanishes, for strong phononic couplings 
the magnetoelastic polarons extend over a much larger phase space and
will dominate the phase diagram of the KLM. 

Immediately below the critical line ($\delta >0$), the system
represents a weakly disordered Griffiths phase. The remaining
polarons occupy a small fraction of the system length but
behave as if they are still in the ordered phase; their
magnetization $\delta^{\beta}$ per unit length is identical
to $M_{0}$ of the weakly ordered phase.\cite{fisher} These 
remaining rare polarons dominate the low-energy physics. Hence, 
crossing the phase transition line disorders the polarons. The 
transition is indeed an order-disorder transition of the magnetic 
polarons. This regime can be viewed as a paramagnet with locally 
ordered ferromagnetic regions. This scenario resembles a two-fluid 
picture, i.e., a polaronic liquid, with intrinsic inhomogeneities 
which involves spin fluctuations and short-range spin correlations. 

\begin{figure} 
\includegraphics[width=5in]{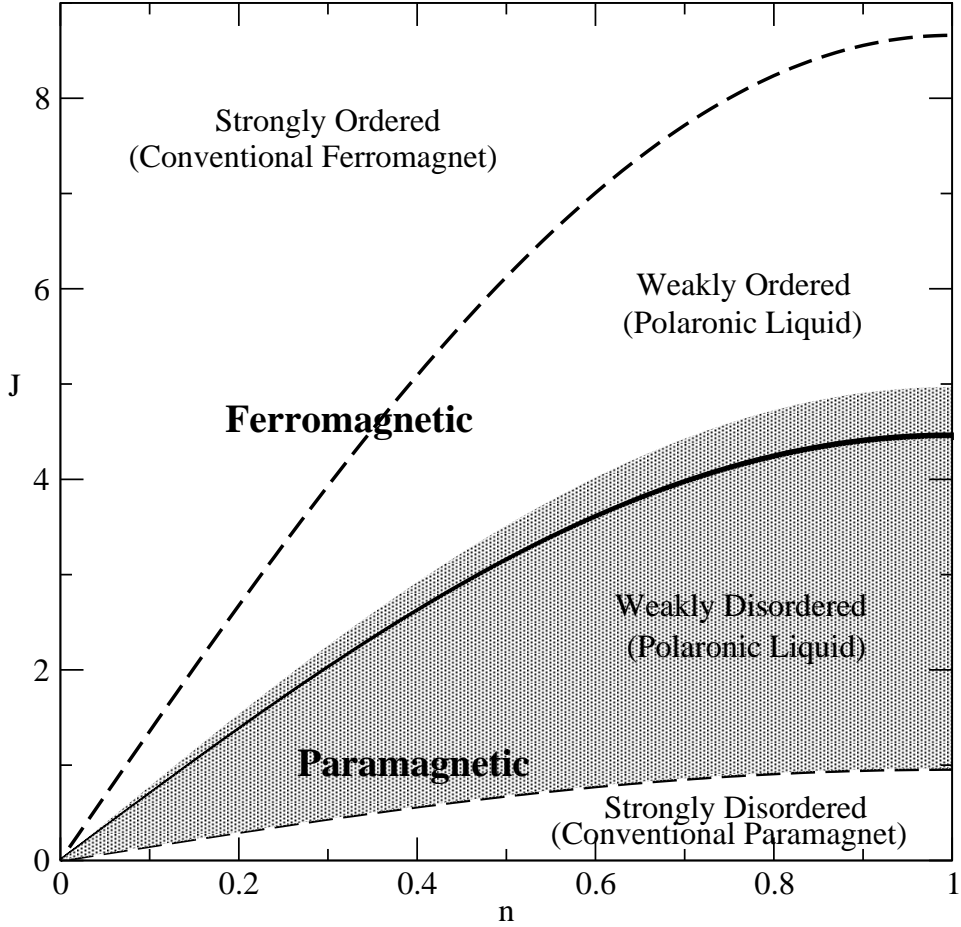} 
\caption{\label{fig:Fig6} The ground-state phase diagram for 
$U = 0$ and $\beta = 5$. The solid line is true phase transition
line from Eq.\ (\ref{jcrit}). The dashed lines separate the conventional 
ferromagnet / paramagnet phases from their weakly disordered 
(polaronic liquid) counterparts. Note the extent of the polaronic 
liquid phases compared to the $U = 0$ and $\beta = 0$ standard KLM 
(shaded region) corresponding to Fig.\ (\ref{fig:Fig3}).} 
\end{figure} 

In this weakly disordered phase, the magnetization 
$M(H) \propto \delta^{\gamma} \{ H^{2\delta}
[\delta \ln(1/H) + {\rm{const.}}] + {\cal{O}}
[H^{4\delta} \delta \ln(1/H)] \}$; thus
$M(H)$ has a power law singularity with a continuously
variable exponent $2\gamma$; as in the weakly ordered
phase the susceptibility is infinite. 

The mean correlation function decays less rapidly than in
the weakly ordered phase, but takes the same form
$(\xi /x)^{5/6} e^{-x/ \xi} \exp[-3/2 (\pi x/\xi)^{1/3}]$ 
for $x \gg \xi = 1/\delta^{2}$. According to
$H_{\rm crit}$, the weakly disordered Griffiths 
phase extends down to $J=0$. However, as the
disorder increases, the third term in $H_{\rm eff}$ is 
no longer negligible. At very low $J$, the last two terms in
$H_{\rm eff}$ will dominate; this corresponds to free spins 
in a field with dominant correlations at $2k_{F}$ of the 
conduction band, and is responsible for the observed peak 
in the localized spin structure factor.\cite{numerics}
This is the strongly disordered conventional 
paramagnetic phase, see Fig.\ \ref{fig:Fig3}. 

As the boundary of this Griffiths phase
is governed by the full $H_{\rm eff}$ from Eq.\ (\ref{heff})  
rather than $J_{\rm crit}$ only, it
will not be effected by $U$ and $\beta$. This
explains why the extent of this Griffiths phase increases
dramatically in the presence of phonons and vanishes for 
large $U$. Thus the phonons indeed will enhance charge 
localization and increase the polaronic effect and act
destructively on the spin ordering. In the presence of 
phonons, these magnetoelastic polarons will dominate a 
larger phase space below and above the ferromagnetic 
transition compared to the standard KLM. This is 
shown explicitly in Fig.\ \ref{fig:Fig6}, where the 
$\beta = 5$ case is compared to the $\beta = 0$, and
$U = 0$ in both cases. It can be seen that the polaronic 
liquid occupies a much larger phase space in the first 
case compared to the standard KLM (shaded region). 
This proves that, at least in the one dimensional 
KLM the local inhomogeneities are enhanced by phonons,
creating magnetoelastic polarons similar to intrinsic
mesoscopic patterns from two-fluid model phenomenologies. 

It is interesting to mention that similar behavior is
expected in higher dimensions also. As shown in 
Ref.\ (\onlinecite{2Dtheory}), the randomness generated through 
local inhomogeneities is the driving force, rather then 
dimensionality, in any real space renormalization approach. 
Thus most of the properties presented previously will survive 
in higher dimensions. This have been confirmed also by
extensive numerical calculations.\cite{2Dnumerics}

An important prediction of our theory is the finite temperature
behavior of the polaronic liquid phase below the transition line, 
the so-called weakly disordered paramagnetic phase. This is the 
region of the phase diagram which is enhanced by phonons and
as such dominated by magnetoelastic polarons, as explained
previously. Here, following closely Ref.\ (\onlinecite{fisher}), 
we can determine the finite temperature susceptibility. At 
small temperatures the weakly coupled 
polarons behave as free spins with momenta $\delta^{\gamma - 1}
\ln (T^{*}/T)$ and the susceptibility will become 
$\chi_{\rm polaron} \approx \delta^{2 \gamma} 
(T/T^{*})^{2 \delta -1} \ln^2 (T/T^{*})$. $T^{*}$ is a temperature 
scale at which the renormalization flow is stopped when the 
temperature fluctuation scale equals the quantum fluctuation

The susceptibility diverges at $T \rightarrow 0$ in the
Griffiths phase as expected \cite{griffiths}. However, 
if $\delta$ increases, e.g., $\delta \approx 1/2$ 
the system moves away from criticality, the susceptibility
at zero temperature will be finite and will no longer be 
dominated by the large rare polarons but rather by the 
more typical smaller ones, becoming a strongly disordered
phase. On the temperature scale, this corresponds 
to increasing the temperature. Thus, at higher 
temperatures the susceptibility will become
$\chi_{\rm polaron} \approx (T/T^{*} - 1) \ln (T/T^{*}) 
= (1 - T/T^{*}) \ln (T^{*}/T)$. Interestingly, 
this is identical to the phenomenological form of the 
susceptibility that has been associated with heavy 
quasiparticles in all heavy fermion compounds.\cite{curro} 
For these compounds $T^{*}$ is the temperature at which the 
heavy quasiparticle liquid start to emerge from a standard 
Kondo lattice behavior. 
Note that in our approach of deriving $\chi_{\rm polaron}$
we moved away from the critical point, in agreement with
recent calculations \cite{hf} that the heavy fermion
systems are in the vicinity off, but not on a Griffiths 
singularity. 

\section{Conclusions}

In summary, we have derived an effective Hamiltonian from a 
one-dimensional Kondo lattice model extended to include effects 
stemming from coupling to the lattice and in the presence of an onsite 
Hubbard term, which accounts for the conduction electron 
Coulomb repulsion. The results are: i) A ferromagnetic 
phase appears at intermediate $\vert J \vert$ due to 
forward scattering by delocalized conduction electrons. 
ii) Ferromagnetism is favoured by the Hubbard term, 
while it is suppressed by the electron-phonon coupling. 
iii) The paramagnetic phase is characterized by 
the coexistence of polaronic regimes with intrinsic 
ferromagnetic order and ordinary conduction electrons. 
iv) In the paramagnetic phase, two time scales compete 
with each other - reminiscent of a two-fluid model -  
and the variability of the critical exponents suggests 
the existence of a Griffiths phase. 

The results are related 
to the small-doping regime of CMR materials which are 
ferromagnets at low temperatures, since here the coupling 
to the phonons has been shown to dominate the 
paramagnetic-ferromagnetic phase transition. 
Note that moving away from this criticality into the 
weakly disordered polaronic liquid phase the results,
in particular the finite temperature susceptibility, 
show behavior characteristic to all heavy fermion 
compounds.\cite{curro}

In regard to the CMR materials, it is interesting to 
point out the discrepancy between infinite 
dimensional calculations and the present one dimensional 
results. Many approximate calculations to model 
CMR \cite{phonontheory} have been made in dynamical 
mean-field theory, which is an infinite dimensional 
approximation and therefore incapable of capturing spatial 
inhomogeneities. In the present work we have approached the 
CMR materials via a one dimensional approximation, but with 
techniques able to describe fluctuations of short-range order. 
Our results show that strong intrinsic spatial inhomogeneities of 
Griffiths type dominate the behaviour of the Kondo lattice. 
Consequently the inhomogeneities exhibit clear 
statistical scaling properties as a function of the proximity
to a quantum (order-disorder) critical point. 
The phonons enhance the inhomogeneities, which, in a good 
approximation behave as a supercritical 
(metastable) phase of a two fluid model.

Even though various bosonization schemes have been used for 
the one-dimensional KLM,\cite{boso,graeme} non of the previous 
approaches took into account phonons. The inclusion of phonon 
degrees of freedom has been shown here to be relevant in creating
local magnetic inhomogeneities. It is important to emphasize
that the properties of the system are controlled by intrinsic 
inhomogeneities. This means that, in a renormalization group 
approach, the dimensionality should not matter.\cite{2Dtheory,2Dnumerics} 
Thus, a similar behaviour is expected in realistic two- and three-dimensions, 
which clearly merit further detailed study. 

\section*{Acknowledgements}

One of the authors (ARB) acknowledges stimulating discussion with 
N. J. Curro concerning the data in Ref.\ \onlinecite{curro}. Work 
at Los Alamos is supported by the US DoE.


\begin{references}

\bibitem{deCMR}C. W. Searle and S. T. Wang, Can. J. Phys. 
{\bf 48}, 2023 (1970); K. Kubo and N. Ohata, J. Phys. Soc. 
Jpn {\bf 63}, 3214 (1972). 

\bibitem{hayden} S. M. Hayden, G. H. Lander, J. Zaretsky, P. J. Brown, C. 
Stassis, P. Metcalf, and J. M. Honig, Phys. Rev. Lett. {\bf 68}, 1061 (1992).
\bibitem{chen}C. H. Chen, S-W. Cheong, and A. S. Cooper, Phys. Rev. Lett.
{\bf 71}, 2461 (1993).

\bibitem{kawano}H. Kawano, R. Kajimoto, H. Yoshizawa, Y. Tomoioka, 
H. Kuwahara, and Y. Tokura, Phys. Rev. Lett. {\bf 78}, 4253 (1997). 

\bibitem{tomioka}Y. Tomioka, A. Asamitsu, H. Kuwahara, and Y. Moritomo, 
Phys. Rev. B {\bf 53}, R1689 (1996).

\bibitem{li}J. Q. Li, Y. Matsui, T. Kimura, and Y. Tokura, 
Phys. Rev. B {\bf 57}, R3205 (1998).

\bibitem{moritomo}Y. Moritomo, A. Nakamura, S. Mori, N. Yamamoto, K. Ohoyama,
and M. Ohashi, Phys. Rev. B {\bf 56}, 14879 (1997).

\bibitem{chen2}C. H. Chen and S-W. Cheong, Phys. Rev. Lett. 
{\bf 76}, 4042 (1996).

\bibitem{wollan}E. O. Wollan and W. C. Koehler, Phys. Rev. {\bf 100}, 545
(1955).

\bibitem{liubao}H. L. Liu, A. L. Cooper, and S-W. Cheong, 
Phys. Rev. Lett. {\bf 81}, 4684 (1998); W. Bao, J. D. Axe, C. H. Chen, 
and S-W. Cheong, Phys. Rev. Lett. {\bf 78}, 543 (1997).

\bibitem{mori}S. Mori, C. H. Chen, and S-W. Cheong, 
Nature {\bf 392}, 473 (1998).

\bibitem{ueharajirac}M. Uehara, S. Mori, C. H. Cheng, and S-W. Cheong, 
Nature {\bf 399}, 560 (1999). 

\bibitem{zener51}C. Zener, Phys. Rev. {\bf 82}, 403 (1951). 

\bibitem{phonontheory}See, for example A. J. Millis, P. B. Littlewood and 
B. I. Shraiman, Phys. Rev. Lett. {\bf 74}, 5144 (1995); 
H. R\"oder, J. Zang and A. R. Bishop, Phys. Rev. Lett. 
{\bf 76}, 1356 (1996); and the references cited therein. 

\bibitem{MC}E. Dagotto, S.Yunoki, A. L. Malvezzi, A. Moreo, J. Hu,
S. Capponi, D. Poilblanc, and N. Furukawa,
Phys. Rev. B{\bf 58}, 6414 (1998); S. Yunoki, J. Hu, A. L. Malvezzi,
A. Moreo, N. Furukawa, and E. Dagotto, Phys. Rev. Lett. {\bf 80}, 845 (1998); 
H. Yi, N. H. Hur, and J. Yu, Phys. Rev. B{\bf 61}, 9501 (2000); 
Y. Motome and N. Furukawa, J. Phys. Soc. Jpn. {\bf 69},3785 (2000); 
W. Koller, A. Pr\"{u}ll, H. G. Evertz, and W. von der Linden, Phys.
Rev. B{\bf 67}, 174418 (2003). 

\bibitem{heavyfermion}See, for example, P. A. Lee, T. M. Rice, J. W.
Serese, L. J. Sham, and J. W. Wilkins, Comments Condens. Matter 
Phys. {\bf 12}, 99 (1986). 

\bibitem{numerics}M. Troyer and D. W\"{u}rtz, Phys. Rev. 
B{\bf 47}, 2886 (1993); H. Tsunetsugu, M. Sigrist, and K. Ueda, 
Phys. Rev. B{\bf 47}, 8345 (1993); S. Moukouri and L. G. Caron, 
Phys. Rev. B{\bf 52}, R15723 (1995); N. Shibata and K. Ueda, 
J. Phys. Cond. Matter. {\bf 11}, R1 (1999); I. P. McCulloch, 
A. Jouzapavicius, A. Rosengren, and M. Gulacsi, Phil. Mag. Lett.
{\bf 81}, 869 (2001), and Phys. Rev. B{\bf 65}, 052410 (2002);
A. Jouzapavicius, I. P. McCulloch, M. Gulacsi and A. Rosengren,
Phil. Mag. B{\bf 82}, 1211 (2002). 

\bibitem{boso}O. Zachar, S. A. Kivelson, and V. J. Emery,
Phys. Rev. Lett. {\bf 77}, 1342 (1996); E. Novais, E. Miranda, 
A. H. Castro Neto, and G. G. Cabrera, Phys. Rev. B{\bf 66}, 174409 (2002). 

\bibitem{graeme}G. Honner and M. Gulacsi, Phys. Rev. Lett.
{\bf 78}, 2180 (1997); Phys. Rev. B{\bf 58}, 2662 (1998). 

\bibitem{yuno1}S. Yunoki and A. Moreo, Phys. Rev. B {\bf 58}, 6403 (1998). 

\bibitem{Martin-Mayor}J. L. Alonso, J.A. Capitan, L.A. Fernandez, 
F. Guinea and V. Martin-Mayor, Phys. Rev. B {\bf 64}, 054408 (2001).

\bibitem{ssh}W. P. Su, J. R. Schrieffer and A. J. Heeger, Phys. Rev.
Lett. {\bf 42}, 1698 (1979); Phys. Rev. B{\bf 22}, 2099 (1980). 

\bibitem{holstein}T. Holstein, Ann. Phys. (N.Y.) {\bf 8}, 
325; 343 (1959).

\bibitem{hirsch}J. E. Hirsch and E. Fradkin, Phys. Rev. Lett. {\bf 49}, 402
(1982); Phys. Rev. B{\bf 27} 4302 (1983); E. Fradkin and J. E. Hirsch 
Phys. Rev. B{\bf 27}, 1680 (1983). 

\bibitem{haldane}F. D. M. Haldane, J. Phys. C{\bf 14}, 2585 (1981). 
For a review, see J. Voit, Rep. Prog. Phys. {\bf 57}, 977 (1994); 
M. Gulacsi, Phil. Mag. B{\bf 76}, 731 (1997); J. von Delft and H. 
Sch{\"o}ller, Ann. Phys. {\bf 4}, 225 (1998). 

\bibitem{detheory}P. W. Anderson and H. Hasegawa, Phys. Rev. 
{\bf 100}, 675 (1955); P. -G. de Gennes, Phys. Rev. {\bf 118}, 
141 (1960). 

\bibitem{Makhankov}See, for example, V. G. Makhankov, 
 {\it Soliton Phenomenology}, Mathematics and Its Applications 
 (Soviet Series) Vol. 33 (Kluwer, Dordrecht, 1989).
 
\bibitem{curro}N. J. Curro, B. L. Young, J. Schmalian, and 
D. Pines, cond-mat/0402179. 

\bibitem{dilute}M. Gulacsi, I. P. McCulloch, A. Jouzapavicius, 
and A. Rosengren, cond-mat/0304351. 

\bibitem{pfeuty}P. Pfeuty, Ann. Phys. (N.Y.) {\bf 57}, 79 (1970).

\bibitem{fisher}D. S. Fisher, Phys. Rev. Lett. {\bf 69},
534 (1992); Phys. Rev. B {\bf 51}, 6411 (1995). 

\bibitem{abrikosov}See, for example, A. A. Abrikosov and
S. I. Moukhin, J. Low Temp. Phys. {\bf 33}, 207 (1978). 

\bibitem{griffiths}R. B. Griffiths, Phys. Rev. Lett.
{\bf 23}, 17 (1969). 

\bibitem{2Dtheory}D. S. Fisher, Physica A{\bf 263}, 222 (1999); 
O. Motrunich, S.-C. Mau, D. A. Huse, and D. S. Fisher, Phys. 
Rev. B{\bf 61}, 1160 (2000); A. A. Middleton and D. S. Fisher,
Phys. Rev. B{\bf 65}, 134411 (2002).

\bibitem{2Dnumerics}Y.-C. Lin, N. Kawashima, F. Igloi, and
H. Rieger, Prog. Theor. Phys. Supp. No. 138, 479 (2000); 
D. Karevski, Y.-C. Lin, H. Rieger, N. Kawashima, and 
F. Igloi, Eur. Phys. Jour. B{\bf 20}, 267 (2001). 

\bibitem{hf}A.J. Millis, D.K. Morr, and J. Schmalian, 
Phys. Rev. Lett. {\bf 87}, 167202 (2001); Phys. Rev.
B{\bf 66}, 174433 (2002); T. Vojta, Phys. 
Rev. Lett. {\bf 90}, 107202 (2003). 

\end{references}
\end{document}